# How Much and What Kind of Nonlocality?
# A Sufficient Condition for Singlet Spin Correlations

*Mehmet Ali Kuntman*
*and*
*Ertan Kuntman[1]*

*Middle East Technical University (METU), Department of Physics, 06531 Ankara*

We give a sufficient condition of nonlocality in order to reproduce singlet spin correlations. For a given pair of hidden variables and measurement directions this condition determines only the product of the outcomes and reproduces statistical correlations for all measurement directions; but fails to give a complete description of sub-systems, provides no means to calculate all joint probabilities and puts no constraints on signaling. In order to complete the model we introduce an additional condition. In this case we observe that the character of the nonlocality changes (outcome independence is violated) and nonlocality applies asymmetrically to the spacelike separated parties. At first sight it seems possible to explain this asymmetry by assuming that the observer that measures first determines his outcome locally. However this assumption has no meaning without adopting a privileged reference frame.

## Introduction: Bell-Mermin Hidden Variable Model for Single Spin

In order to show that John von Neumann's theorem[1] against the possibility of hidden variable theories is based on an unjustified assumption, J.S. Bell[2] gives a simple hidden variable model for spin ½. In this model, hypothetical "dispersion free" states are regarded as ensembles of states further specified by additional variables such that values of these variables together with the quantum mechanical state vector determine precisely the results of all possible measurements that can be performed on individual systems. A somewhat simplified version of the rule which specifies the results of measurements is given by David Mermin[3]. According to this rule,

If $(\langle \sigma \rangle_\psi + \lambda) \cdot \mathbf{a} > 0$ result of the measurement is $+1$     (1),

If $(\langle \sigma \rangle_\psi + \lambda) \cdot \mathbf{a} < 0$ result of the measurement is $-1$     (2),

where $\langle \sigma \rangle_\psi$ is the expectation value of the spin vector $\sigma$, $\lambda$ is a random unit vector (which plays the role of the hidden variable) and $\mathbf{a}$ is also a unit vector which specifies the direction of measurement. In short, sign of the inner product $(\langle \sigma \rangle_\psi + \lambda) \cdot \mathbf{a}$ determines the outcomes (*X*) of individual measurements and it can be shown that integral of the sign of this value function on a unit sphere reproduces the quantum mechanical result $\langle \sigma \cdot \mathbf{a} \rangle_\psi = \langle \sigma \rangle_\psi \cdot \mathbf{a}$, i.e.,

---

[1] E-mail: e160442@metu.edu.tr

$$\langle X \rangle = \int \mathrm{sgn}\left[\left(\langle \boldsymbol{\sigma}\rangle_\psi + \boldsymbol{\lambda}\right)\cdot \mathbf{a}\right]\frac{d\Omega}{4\pi} = \langle \boldsymbol{\sigma}\rangle_\psi \cdot \mathbf{a} \quad (3).$$

In general, state of the system can be represented by a density matrix $\rho = \frac{1}{2}(\mathbf{I} + \mathbf{P}\cdot\boldsymbol{\sigma})$, where $\mathbf{P} = \langle \boldsymbol{\sigma}\rangle_\rho$ and $\mathbf{I}$ is the unit matrix, then Eq. (3) reads,

$$\langle X \rangle = \int \mathrm{sgn}[(\mathbf{P} + \boldsymbol{\lambda})\cdot \mathbf{a}]\frac{d\Omega}{4\pi} = \mathbf{P}\cdot \mathbf{a} \quad (4)$$

In a special case for $\mathbf{P} = 0$, outcomes $+1$ and $-1$ occur with equal probabilities ½ and for all directions of $\mathbf{a}$ gives,

$$\langle X \rangle = \int \mathrm{sgn}(\boldsymbol{\lambda}\cdot \mathbf{a})\frac{d\Omega}{4\pi} = 0 \quad (5).$$

This hidden variable model proposed by Bell successfully reproduces all quantum mechanical results for single spin. On the other hand, an important feature of quantum theory lies in the statistical correlations between two parties of entangled systems which are now considered as a substantial part of physical reality. It is also shown by John Bell[4] that such correlations cannot be explained by descriptions based only on local properties of sub-systems and local causes. As is well known, in case of correlated spin of EPR/B[5][6], i.e., for the singlet state of two spin ½,

$$|\psi\rangle = \frac{1}{\sqrt{2}}(|\uparrow\downarrow\rangle - |\downarrow\uparrow\rangle) \quad (6),$$

quantum mechanical correlations are stronger than that of any local hidden variable theory. In short, non-classical correlations cannot be modelled in local hidden variable theories.

## Sufficient Condition for Singlet Spin Correlations

We introduce a simple model which reproduces the singlet spin correlations but fails to give a complete description of sub-systems.

We consider two spatially separated observers $A$ and $B$. They share two random variables $\lambda_1$ and $\lambda_2$ which are real three dimensional unit vectors distributed uniformly over the unit sphere. Observers $A$ and $B$ measure the spin of sub-systems along the directions of the unit vectors $\mathbf{a}$ and $\mathbf{b}$ and in each case obtain the results $X$ and $Y \in \{+1, -1\}$ respectively.

Now we suggest a simple relation such that

$$XY = \mathrm{sgn}(\lambda_1 \cdot \lambda_2 - \mathbf{a}\cdot\mathbf{b}) \quad (7).$$

This relation determines only the *product* of the two parties' outcomes. If $\lambda_1 \cdot \lambda_2 > \mathbf{a}\cdot\mathbf{b}$, $XY = +1$, if not $-1$ and this is enough to calculate the joint expectation value $\langle XY \rangle$, where,

$$\langle XY \rangle = \int \mathrm{sgn}(\lambda_1 \cdot \lambda_2 - \mathbf{a}\cdot\mathbf{b})\frac{d\Omega_1}{4\pi}\frac{d\Omega_2}{4\pi} \quad (8)$$

In order to evaluate the joint expectation value we fix $\lambda_1$ and integrate over $\lambda_2$ first. If $\lambda_2$ resides in the shaded region shown in Figure 1, $\lambda_1 \cdot \lambda_2 > \mathbf{a} \cdot \mathbf{b}$, $\text{sgn}(\lambda_1 \cdot \lambda_2 - \mathbf{a} \cdot \mathbf{b}) = +1$, otherwise $-1$.

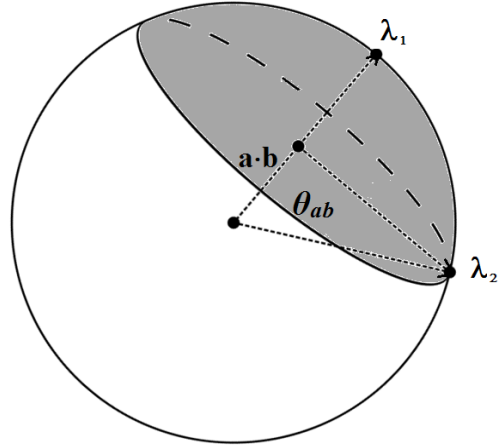

**Figure 1**. We fix $\lambda_1$ and integrate over $\lambda_2$. Vectors $\mathbf{a}$ and $\mathbf{b}$ are not shown. All $\lambda_2$ satisfying $\lambda_1 \cdot \lambda_2 > \mathbf{a} \cdot \mathbf{b}$ reside in the shaded region.

Area of the shaded region is equal to the solid angle subtended by the spherical cap. Therefore,

$$\Omega_2(+) = 2\pi(1 - \cos(\theta_{ab})) \qquad (9).$$

Similarly for the unshaded region,

$$\Omega_2(-) = 2\pi(1 + \cos(\theta_{ab})) \qquad (10),$$

where $\cos(\theta_{ab}) = \mathbf{a} \cdot \mathbf{b}$.

$\Omega_2(+) - \Omega_2(-) = -4\pi \cos(\theta_{ab})$. After normalization with $1/4\pi$ the integral reduces to a simple integral over $\lambda_1$,

$$\langle XY \rangle = \int (-\mathbf{a} \cdot \mathbf{b}) \frac{d\Omega_1}{4\pi} \qquad (11).$$

Hence we obtain,

$$\langle XY \rangle = -\mathbf{a} \cdot \mathbf{b} \qquad (12).$$

This is the quantum mechanical correlation expected for singlet spin ½ and we like to introduce the relation $XY = \text{sgn}(\lambda_1 \cdot \lambda_2 - \mathbf{a} \cdot \mathbf{b})$ as a sufficient condition for a hidden variable model of singlet spin to reproduce quantum mechanical correlations. It is obvious that this relation depends only on the polar angle $\theta_{ab}$ between the measurement settings of A and B, but it does not depend on the *direction* of measurements completely.

This model is nonlocal in the sense that product of the outcomes X and Y depends on the settings of the both parties. On the other hand this model is just nonlocal enough to reproduce singlet spin correlations and puts no constraints on signaling. Furthermore it does not give a complete description of the sub-systems and it does not provide means to calculate marginal probabilities. It permits us to calculate only sum of the joint probabilities $P(+,+) + P(-,-)$ and $P(+,-) + P(-,+)$, where $P(+,+)$ denotes $P(X = +1, Y = +1)$, etc.

## Complete Model Which Gives All Joint Probabilities

We have shown that the sufficient condition given by Eq.(7) reproduces quantum mechanical correlations, but fails to give all joint probabilities. In order to determine the joint probabilities we should impose an additional condition. As a specific example let us suppose that outcomes of $A$ depend only on local variables such that,

$$X = \text{sgn}(\mathbf{a} \cdot \boldsymbol{\lambda}_1) \quad (13).$$

Then from Eq.(7) we have,

$$\text{sgn}(\mathbf{a} \cdot \boldsymbol{\lambda}_1) Y = \text{sgn}(\boldsymbol{\lambda}_1 \cdot \boldsymbol{\lambda}_2 - \mathbf{a} \cdot \mathbf{b}) \quad (14),$$

or equivalently,

$$Y = \text{sgn}(\boldsymbol{\lambda}_1 \cdot \boldsymbol{\lambda}_2 - \mathbf{a} \cdot \mathbf{b}) \, \text{sgn}(\mathbf{a} \cdot \boldsymbol{\lambda}_1) \quad (15)$$

and the expectation value of $Y$ can be calculated as follows:

$$\langle Y \rangle = \int \text{sgn}(\boldsymbol{\lambda}_1 \cdot \boldsymbol{\lambda}_2 - \mathbf{a} \cdot \mathbf{b}) \, \text{sgn}(\mathbf{a} \cdot \boldsymbol{\lambda}_1) \frac{d\Omega_1}{4\pi} \frac{d\Omega_2}{4\pi} \quad (16).$$

This integral can be evaluated in a similar way described above: We fix $\boldsymbol{\lambda}_1$ integrate over $\boldsymbol{\lambda}_2$ and obtain,

$$\langle Y \rangle = \int (-\mathbf{a} \cdot \mathbf{b}) \, \text{sgn}(\mathbf{a} \cdot \boldsymbol{\lambda}_1) \frac{d\Omega_1}{4\pi} \quad (17).$$

On the other hand, according to Eq. (5), $\int \text{sgn}(\mathbf{a} \cdot \boldsymbol{\lambda}_1) \frac{d\Omega_1}{4\pi} = \langle X \rangle = 0$, therefore,

$$\langle Y \rangle = \langle X \rangle = 0 \quad (18).$$

Now we are in a position to calculate all joint probabilities,

$$P(+,+) = P(-,-) = \frac{1}{4}(1 - \cos(\theta_{ab})) \; ; \; P(+,-) = P(-,+) = \frac{1}{4}(1 + \cos(\theta_{ab})) \quad (19).$$

## Discussion

When we introduce Eq. (13) we get a complete nonlocal hidden variable model which reproduces all statistical predictions of quantum mechanics. Our first observation is that the complete model is no more *statistically* parameter dependent and prohibits superluminal signaling, i.e., the rather faint parameter dependence which appears in Eq.(7) does not manifest itself in the conditional probabilities. On the other hand, character of the nonlocality is now changed and additional nonlocality is asymmetrically distributed between two parties: while outcomes $X$ depend only on local variables, outcomes $Y$ depend on both settings of $A$ and outcomes $X$. It is easy to check that this outcome dependence manifests itself also in the conditional probabilities, as is the case in quantum mechanics[7].

Right hand side of the relation $XY = \text{sgn}(\lambda_1 \cdot \lambda_2 - \mathbf{a} \cdot \mathbf{b})$ is not factorizable, any extension of this model will end up with an asymmetric nonlocal model. As noted by Ghirardi and Romano[8] Bell's original nonlocal model is also asymmetric in this sense and Toner and Bacon[9] base their model on an asymmetric protocol.

Is asymmetry an inevitable issue for all nonlocal models? Is it possible to distribute nonlocality symmetrically between two parties? What are the implications of this asymmetry? We are not in a position to answer these questions and we don't want to be hasty to explain this asymmetry with making reference to the observer that measures first: which may recall the notion of privileged reference frame. Anyway we want to believe that the relation $XY = \text{sgn}(\lambda_1 \cdot \lambda_2 - \mathbf{a} \cdot \mathbf{b})$ provides a *sufficient* condition for hidden variable models in order to reproduce quantum mechanical correlations and draws a borderline between quantum mechanics and local hidden variable theories.